\documentclass[a4paper,fleqn,usenatbib]{mnras}

\usepackage{mathptmx}

\usepackage[T1]{fontenc}
\usepackage{ae,aecompl}

\usepackage{graphicx}	
\usepackage{amsmath}	
\usepackage{amssymb}	

\usepackage{times}
\usepackage{longtable}
\usepackage{multirow}
\usepackage{rotating}
\usepackage{color}
\usepackage{booktabs}
\usepackage{epstopdf}
\usepackage{bm}
\usepackage{indentfirst}
\usepackage{booktabs}
\usepackage{indentfirst}
\usepackage{threeparttable}
\usepackage{subfigure}
\usepackage{mathrsfs}
\usepackage{url}

\title{Polarisation of submillimetre lines from interstellar medium}

\author[Zhang H., Yan H.]{
Heshou Zhang,$^{1,2}$
Huirong Yan,$^{1,2}$\thanks{E-mail: hyan@mail.desy.de}
\\
$^{1}$Deutsches Elektronen-Synchrotron DESY, Platanenallee 6, D-15738 Zeuthen, Germany\\
$^{2}$Institut f$\ddot{u}$r Physik und Astronomie, Universit$\ddot{a}$t Potsdam, Haus 28, Karl-Liebknecht-Str. 24/25, D-14476 Potsdam, Germany
}

\pubyear{2017}

\begin{document}
\label{firstpage}
\pagerange{\pageref{firstpage}--\pageref{lastpage}}
\maketitle

\begin{abstract}
Magnetic fields play important roles in many astrophysical processes. However, there is no universal diagnostic for the magnetic fields in the interstellar medium (ISM) and each magnetic tracer has its limitation. Any new detection method is thus valuable. Theoretical studies have shown that submillimetre fine-structure lines are polarised due to atomic alignment by Ultraviolet (UV) photon-excitation, which opens up a new avenue to probe interstellar magnetic fields. We will, for the first time, perform synthetic observations on the simulated three-dimensional ISM to demonstrate the measurability of the polarisation of submillimetre atomic lines. The maximum polarisation for different absorption and emission lines expected from various sources, including Star-Forming Regions (SFRs) are provided. Our results demonstrate that the polarisation of submillimetre atomic lines is a powerful magnetic tracer and add great value to the observational studies of the submilimetre astronomy.
\end{abstract}

\begin{keywords}
submillimetre: ISM -- turbulence -- ISM: magnetic fields -- (ISM:) HII regions -- (ISM:) photodissociation region (PDR) -- polarisation
\end{keywords}



\section{Introduction}

Submillimetre astronomy is an indispensable window for the study of the universe. Submillimetre spectroscopy plays a crucial role in understanding the processes such as galaxy evolution \citep{Sparke2000}, the interstellar medium (ISM), e.g., molecular clouds \citep{Stutzki88}, Photon-Dissociation Regions (PDRs) \citep{Hollenbach1999}, etc. In particular, submillimetre polarisation arising from dust alignment is one of the major magnetic tracers with promising analytical and observational progress (see, e.g., the review by \citealt{ALVgrain2015}). Uncertainties with grain alignment do exist though due to the unknown shape and compositions of the grains. In view of the fact that limitations exist in all the magnetic diagnostics that are currently applied to the observation of magnetic field in the ISM, the exploration with independent techniques would be important and complementary to current methods.

Theoretical works have shown that the polarisation of atomic lines in the submillimetre band can be used to trace magnetic fields in the ISM due to the physical effect of atomic alignment\footnote{For simplicity, the term 'atom' represents atoms and ions. The term 'alignment' here refers to the direction of the angular momenta of the atoms (see, e.g., the review by \citealt{YL15}).}\citep{YLfine,YLhyf,YLhanle,SY2013,ZYD15,ZYR2016}. Submillimetre atomic lines result from the fine-structure transitions between levels in the ground state of atoms. In the diffuse ISM, the photon-excitation is the dominant mechanism to produce submillimetre atomic lines: the massive stars or clusters nearby the ISM radiate Ultraviolet (UV) photons which illuminate the ISM by pumping the atoms from the ground state to the excited states; the atoms end up on different fine-structure levels in the ground state due to the spontaneous emission from the excited states. The radiation source provides a typical anisotropic radiation field which aligns the atoms in the medium. The submillimetre lines emitted and absorbed from the aligned medium are polarised. The alignment is altered according to the direction of the magnetic field due to the fast magnetic precession, which is termed as magnetic realignment. The resultant polarisation of the fine-structure lines thereby reveals interstellar magnetic fields. The magnetic strength in the ISM is generally weak ($\sim\mu G$), which means only ground state alignment occurs (see \citealt{YL12}). In a totally different regime than the ISM\footnote{The solar magnetic field, which is generally stronger than 1 Gauss, resides in Hanle or Zeeman regimes.}, solar physicists have been employing spectral polarimetry to study the solar magnetism (see, e.g., \citealt{Landi-DeglInnocenti:1983mi,Landi-DeglInnocenti:1984kl,Landi-DeglInnocenti:1998pi,Stenflo:1997cq}). Current facilities already have the capability to cover the submillimetre band for the spectral polarimetry observation (see, e.g. \citealt{Risacher16}). Nevertheless, the following questions should be addressed before the observation: the measurability for the polarisation of submillimetre atomic lines and what information of the interstellar magnetic fields could be revealed. Through synthetic observations on numerical simulated ISM, the purposes of this paper are to answer the afore mentioned questions and demonstrate the value of submillimetre spectropolarimetry as a magnetic tracer.

\section{Submillimetre spectropolarimetry in the ISM}

\begin{figure*}
\centering
\subfigure[]{
\includegraphics[width=0.96\columnwidth]{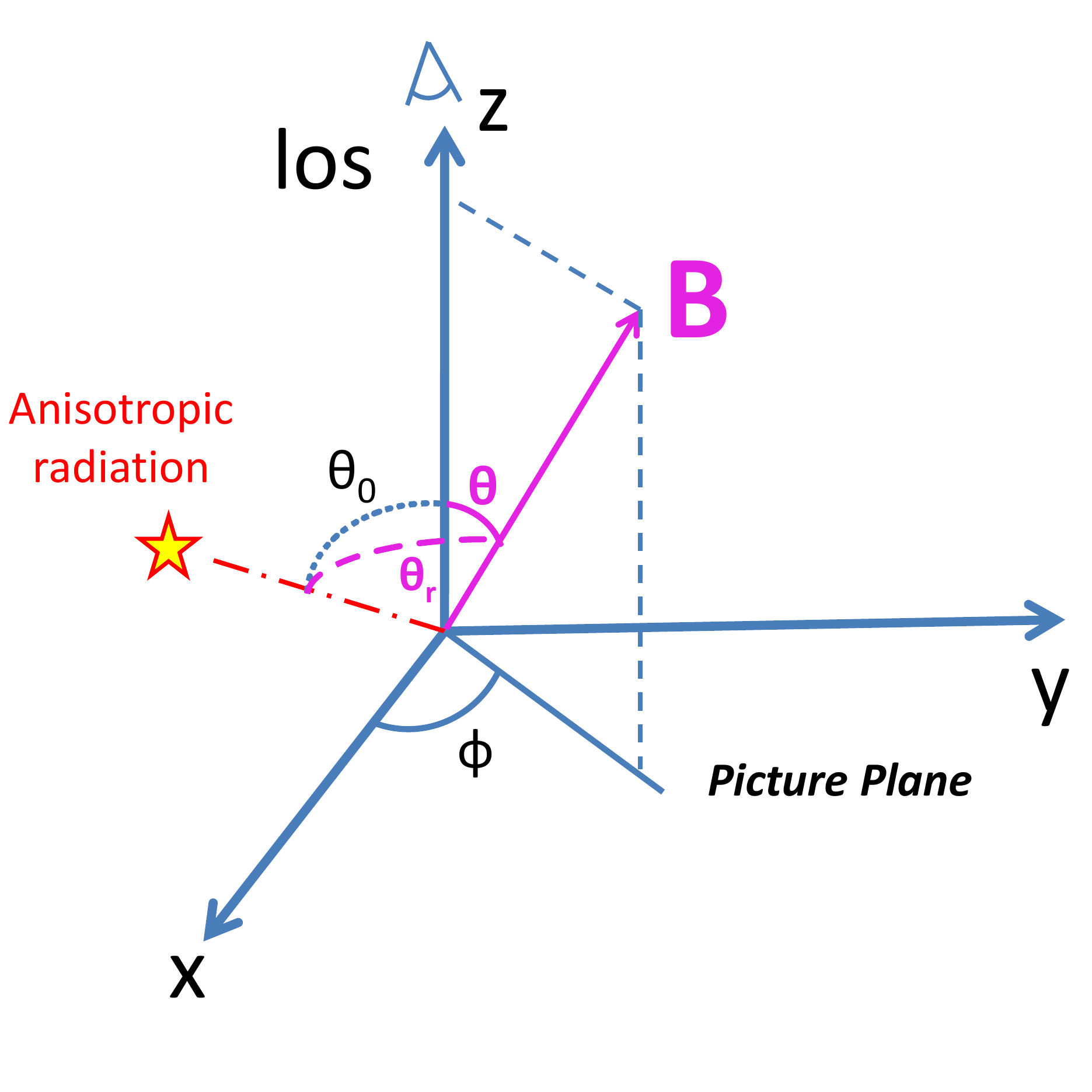}\label{theomap}}
\subfigure[]{
\includegraphics[width=0.96\columnwidth]{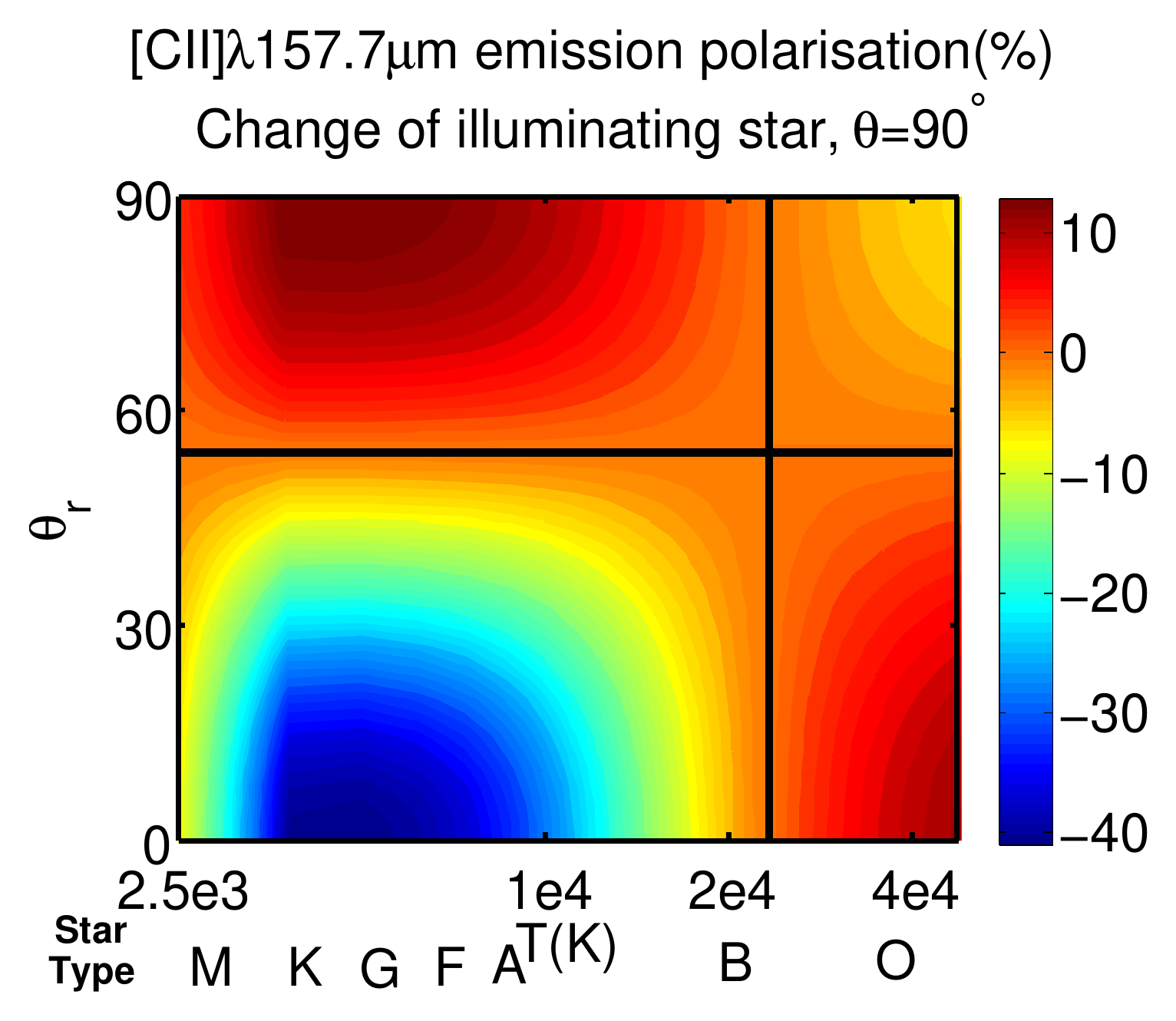}\label{C2pt2}}
\caption{(a) $xyz-$coordinate system with the line of sight along the $z-$axis, the $xy-$plane is the picture plane. The angle between magnetic field and line-of-sight is $\theta$. The angle between the incidental radiation and magnetic field is $\theta_r$. $\theta_0$ is the angle between line of sight and incidental radiation; (b) Polarisation of [C\,{\sc ii}]$\lambda157\mu m$ emission line with different types of pumping source at $\theta_0=90^{\circ}$. The type of radiation source is marked on the $x-$axis at the corresponding temperature. The criteria where the induced polarisation equals zero are marked with dark solid-lines.}\label{Tchange}
\end{figure*}

Atomic lines in the submillimetre band are magnetic dipole transitions between the fine-structure levels in the ground state of the atoms. Hence, the decay rate for the atoms from the fine-structure levels in the ground state is the magnetic dipole transition rate $A_m$, which, in ISM, is much lower than magnetic precession rate $\nu_L$. Thus, the alignment of atomic angular momentum can happen on both the upper and lower fine-structure levels of the magnetic dipole transition within the ground state. Therefore, both submillimetre absorption and emission atomic lines can be polarised (see \citealt{YL12} for details). The spectral polarisation is altered depending on the direction of magnetic field, and the Stokes parameters of the spectral lines, [$I,Q,U,V$], are modulated accordingly (\citealt{Landi-DeglInnocenti:1984kl}, see also \citealt{YLfine}). The circular-polarisation $V-$channel does not exist since only the unpolarised background radiation is considered in this paper\footnote{Otherwise, circular polarisation appears in absorption lines if the alignment direction differs from the direction of polarised background radiation (see \citealt{YLhyf}).}. The polarisation of submillimetre fine-structure lines from level $J_1$ to level $J_2$ in the ground state can be obtained based on results in \citet{YLfine}:
\begin{equation}\label{pemeq}
P=\frac{-3\sqrt{2}\omega^{2}_{J_1J_2}\sigma^2_0(J_1,\theta_r)\sin^2\theta}{4+\sqrt{2}\omega^{2}_{J_1J_2}\sigma^2_0(J_1,\theta_r)(2-3\sin^2\theta)}.
\end{equation}
The angles $\theta_r,\theta$ are defined in Fig.~\ref{theomap}. The alignment parameter $\sigma^2_0\equiv\rho^2_0/\rho^0_0$, in which $\rho^{0,2}_0$ are the irreducible density matrices for the atoms\footnote{For example, the irreducible density matrix tensor for $J/F=1$ is $\rho^2_0=[\rho(1,1)-2\rho(1,0)+\rho(1,-1)]$ (\citealt{Fano1957,Yakonov1965,BommierS1978}).}. The quantity $\omega^2_{J_1J_2}\equiv\{1,1,2;J_1,J_1,J_2\}/\{1,1,0;J_1,J_1,J_2\}$ is related to the atomic structure, where the matrices with $'\{$ $\}'$ are $6-j$ symbols that demonstrate the angular momentum coupling.

The polarisation is measured from the projection of the magnetic field on the picture plane. Thus the sign of the polarisation reveals the direction of the polarisation: '$+$' means parallel to the magnetic field whereas '$-$' means perpendicular. As demonstrated in Fig.~\ref{C2pt2}, the polarisation is flipped from parallel to perpendicular at the flipping criteria $\theta_r=54.7^{\circ}$ (Van Vleck angle, see \citealt{Vleck25,House74}). When calculating the polarisation of the atomic lines, all the excitations from the ground state to multiple excited states should be included. The weight ratio of the transitions to different excited states varies with the type of radiation source. C\,{\sc ii}, for example, has two excited states: $2S_{\frac{1}{2}}$ and $2D_{\frac{3}{2},\frac{5}{2}}$. The wavelength for the transition from the ground state $2P^{\circ}$ to the excited state $2S$ is $1034\mbox{\AA}$ and for the transition to $2D$ is $1334\mbox{\AA}$. As demonstrated in Fig.~\ref{C2pt2}, the photon-excitation for the magnetic dipole transition [C\,{\sc ii}]$\lambda157\mu m$ is dominated by $2P^{\circ}\rightarrow 2D$ for the pumping stars with lower temperature and by $2P^{\circ}\rightarrow 2S$ for higher temperature. The competition between C\,{\sc ii}$\lambda1034\mbox{\AA}$ and C\,{\sc ii}$\lambda1334\mbox{\AA}$ is balanced at $T=2.32\times10^4 K$, where no polarisation is induced for [C\,{\sc ii}]$\lambda157\mu m$. Furthermore, the photon-excitation can be stronger if the radiation source is a star with higher temperature (with younger age), a larger radius (giant stars such as AGB stars), or a cluster with a larger amount of stars. As demonstrated in \citet{YLfine}, the applicable regime to study magnetic fields with spectral polarimetry would be broader with a stronger radiation field.

The submillimetre emission atomic lines are commonly observed in Star Forming Regions (SFRs), where young O,B-type stars are the dominant optical pumping source. The maximum polarisations of submillimetre emission lines from SFRs are presented in Table 1\footnote{It is important for the readers to note that the results in this table are not conflicted with Table in the review \citep{YL12}. The table in that review demonstrated the polarisation induced by Interstellar Radiation Field (ISRF).}. As demonstrated in Table 1, the polarisation of submillimetre emission atomic lines is quite substantial for many elements in SFRs.

\begin{table}
\centering
\caption{MAXIMUM POLARISATION FOR SUBMILLIMETRE EMISSION LINES}\label{subemretable}
\begin{tabular}{|c|c|c|c|}
\hline \hline
Species & Transition & Wavelength & max(P) \\
\hline
[C\,{\sc i}] & $3P_{1}\rightarrow 3P_{0}$ & $610\mu m$ & $21\%^{a}$\\
\hline
[C\,{\sc i}] & $3P_{2}\rightarrow 3P_{1}$ & $370\mu m$ & $18\%^{b}$\\
\hline
[C\,{\sc ii}] & $2P^{\circ}_{3/2}\rightarrow 2P^{\circ}_{1/2}$ & $157.7\mu m$ & $28.5\%^{a}$ \\
\hline
[O\,{\sc i}] & $3P_{1}\rightarrow 3P_{2}$ & $63.2\mu m$ & $4.2\%^{a}$ \\
\hline
[Si\,{\sc i}] & $3P_{1}\rightarrow 3P_{0}$ & $129.7\mu m$ & $20\%^{a}$\\
\hline
[Si\,{\sc i}] & $3P_{2}\rightarrow 3P_{1}$ & $68.5\mu m$ & $18\%^{b}$\\
\hline
[Si\,{\sc ii}] & $2P^{\circ}_{3/2}\rightarrow 2P^{\circ}_{1/2}$ & $34.8\mu m$ & $12.6\%^{b}$ \\
\hline
[S\,{\sc i}] & $3P_{1}\rightarrow 3P_{2}$ & $25.2\mu m$ & $3.2\%^{a}$\\
\hline
[Fe\,{\sc ii}] & $a6D_{7/2}\rightarrow a6D_{9/2}$ & $26.0\mu m$ & $4.9\%^{a}$\\
\hline
\end{tabular}\\
\begin{tablenotes}
      \small
      \item Note: The table considers SFRs with young O,B-type stars $T\sim$[$1\times10^4K,5\times10^4K$]. The sources inducing the maximum polarisation are marked correspondingly. Simulation shows that most of the lines can reach at least $80\%$ of the maximum polarisation with all types of illuminating stars considered. The atomic emission lines are unpolarised if the upper level angular momentum $J_u=0,1/2$.
      \item (a) Maximum at $1\times10^4K$ B-type stars.
      \item (b) Maximum at $5\times10^4K$ O-type stars.
\end{tablenotes}
\end{table}

The submillimetre absorption atomic lines represent either the self-absorption of nebulae, or the absorption by the medium that resides on the line of sight. The maximum polarisation of submillimetre absorption lines with unpolarised background spectra and the corresponding pumping source are presented in Table 2. Note that the solar temperature ($T\simeq6\times10^3K$) is close to the temperature of the maximum-polarisation pumping source for most of the elements listed. Hence, an easy target to apply the polarisation of submillimetre fine-structure absorption lines as a magnetic tracer could be the magnetic field in the diffuse medium in solar system (e.g., comet magnetic fields or magnetic tail of planets).

\begin{table}
\centering
\caption{MAXIMUM POLARISATION FOR SUBMILLIMETRE ABSORPTION LINES}\label{subabretable}
\begin{tabular}{|c|c|c|c|}
\hline \hline
Species & Transition & Wavelength & max($P/\tau$) \\
\hline
[C\,{\sc i}] & $3P_{1}\rightarrow 3P_{2}$ & $370\mu m$ & $2\%^{a}$\\
\hline
[O\,{\sc i}] & $3P_{2}\rightarrow 3P_{1}$ & $63.2\mu m$ & $30.8\%^{b}$ \\
\hline
[O\,{\sc i}] & $3P_{1}\rightarrow 3P_{0}$ & $145.5\mu m$ & $49.1\%^{c}$ \\
\hline
[S\,{\sc i}] & $3P_{2}\rightarrow 3P_{1}$ & $25.2\mu m$ & $30.1\%^{d}$\\
\hline
[S\,{\sc i}] & $3P_{1}\rightarrow 3P_{0}$ & $56.3\mu m$ & $45.2\%^{e}$\\
\hline
[Si\,{\sc i}] & $3P_{1}\rightarrow 3P_{2}$ & $370\mu m$ & $2\%^{a}$\\
\hline
[Fe\,{\sc ii}] & $a6D_{9/2}\rightarrow a6D_{7/2}$ & $26.0\mu m$ & $9.9\%^{f}$\\
\hline
\end{tabular}\\
\begin{tablenotes}
      \small
      \item Note: The table considers all different types of illuminating star with temperature range [$2.5\times10^3K,5\times10^4K$]. The environments that induce the maximum polarisation are marked. Simulation shows that most of the lines can reach at least $80\%$ of the maximum polarisation in any environment for strong pumping. The atomic absorption lines are unpolarised if the lower level angular momentum $J_l=0,1/2$.
      \item (a) Almost no variation of maximum polarisation with different sources.
      \item (b) Maximum at $5\times10^3K$ K-type stars.
      \item (c) Maximum at $6\times10^3K$ F,G-type stars (e.g., the Sun).
      \item (d) Maximum at $3.7\times10^3K$ M-type stars, AGB stars, etc.
      \item (e) Maximum at $4.1\times10^3K$ K-type stars.
      \item (f) Maximum at $1\times10^4K$ B,A-type stars.
\end{tablenotes}
\end{table}

\section{Synthetic Observations}

\begin{figure*}
\centering
\subfigure[]{
\includegraphics[width=0.96\columnwidth]{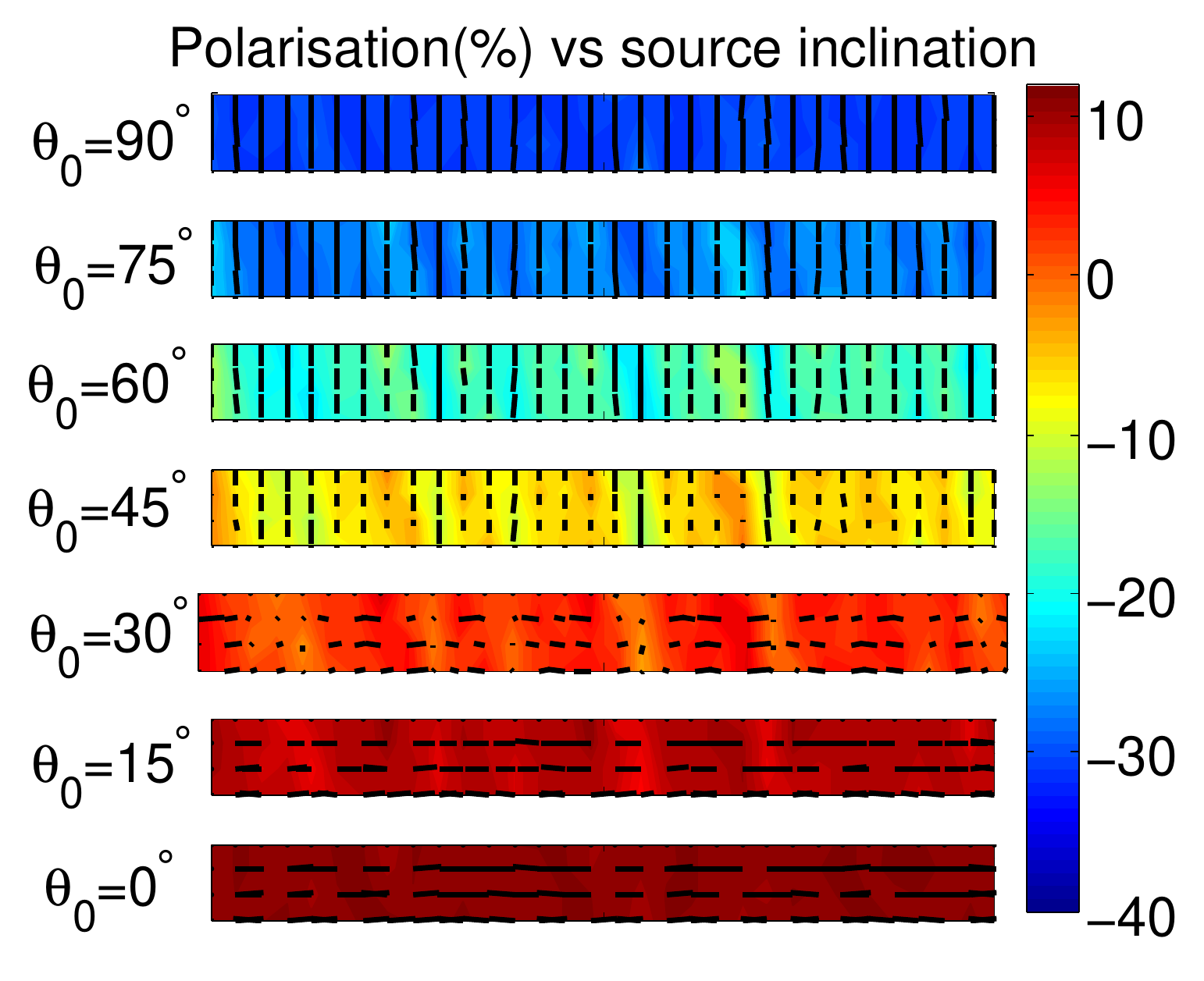}\label{polar_rinc}}
\subfigure[]{
\includegraphics[width=0.96\columnwidth]{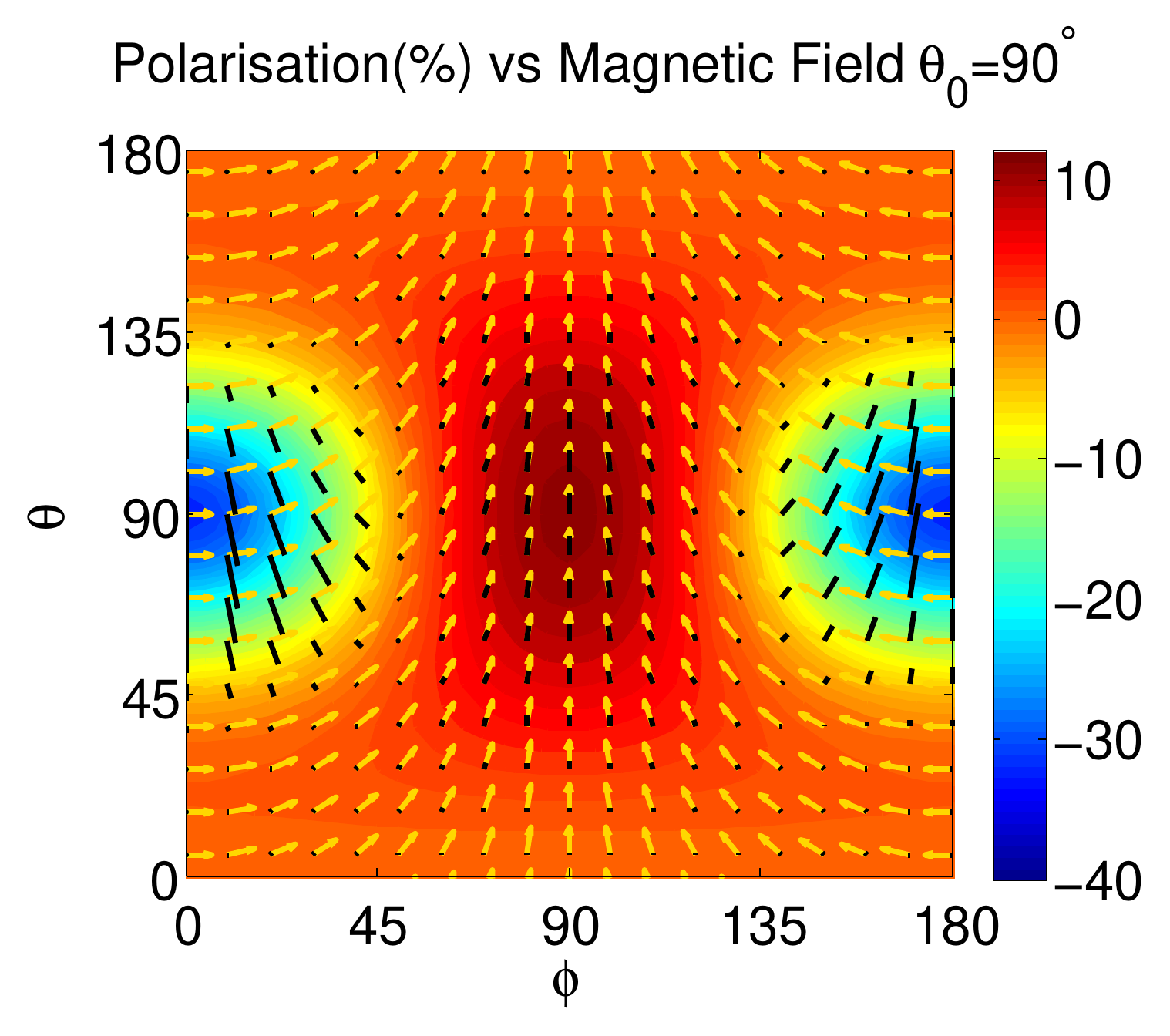}\label{polar_90tht0}}
\caption{(a) [C\,{\sc ii}]$\lambda157\mu m$ polarisation maps for a bar-shaped PDR with different inclination angle of the radiation source $\theta_0$. The mean magnetic direction is along the bar; (b) Polarisation of [C\,{\sc ii}]$\lambda157\mu m$ emission line on the picture plane with different magnetic direction for $\theta_0=90^{\circ}$ with a B9-type pumping star ($T=1\times10^4K$). The black bars are the direction of polarisation expected. The degree of polarisation is marked by the length of bars and the background colour. The magnetic magnetic field on the picture plane is marked as orange arrows.}\label{pomapPDR}
\end{figure*}

Magnetic dipole transitions within the ground state in the diffuse ISM are mainly induced by photon-excitation. Collision excitation can be ignored in the diffuse ISM, whose density is below the critical density \citep{Draine2011}. Therefore, polarisation of submillimetre fine-structure lines is a perfect magnetic tracer in the diffuse ISM. The [C\,{\sc ii}]$\lambda157\mu m$ emission line is selected as the example spectral line in our simulation since $\rm{C}^+$ is one of the most important and most common species in the ISM. Numerical simulations in this paper are performed with the PENCIL-code\footnote{See \url{https://code.google.com/archive/p/pencil-code/} for details.}. First, a bar-shaped PDR with the mean field direction along the bar is considered. Fig.~\ref{polar_rinc} are the maps of polarisation on the picture plane for the radiation source positioned with different line-of-sight angle $\theta_0$. The magnetic component on the picture plane is along the long-axis of the rectangular and the direction of the induced polarisation is either parallel or perpendicular to it. Therefore, the magnetic component on the picture plane could be indicated by the dominant direction of the polarisation with a $90^{\circ}$-degeneracy, which is independent of the direction of incidental radiation. Furthermore, the magnetic direction is varied in the whole space and the resultant polarisation is compared with the magnetic component on the picture plane. As shown in Fig.~\ref{polar_90tht0}, 2-dimensional (2D) polarisation direction readily reveals the projection of magnetic field on the picture plane with a $90^{\circ}$-degeneracy. Furthermore, 3D magnetic direction can be deduced given enough priori information, such as the detection of the polarisation of multiple ($\ge2$) submillimetre lines at the same spot. In addition, the maximum polarisation for the 'parallel' and 'perpendicular' case are different. As demonstrated in Fig.~\ref{pomapPDR}, the 'parallel' case has a maximum of $12\%$ polarisation whereas the 'perpendicular' case is very likely to produce more than $20\%$ polarisation. Hence, the measurement of the degree of polarisation will help break the $90^{\circ}-$degeneracy and thus provide us with an accurate measurement of the magnetic fields.

\begin{figure*}
\centering
 \subfigure[]{
\includegraphics[width=0.96\columnwidth]{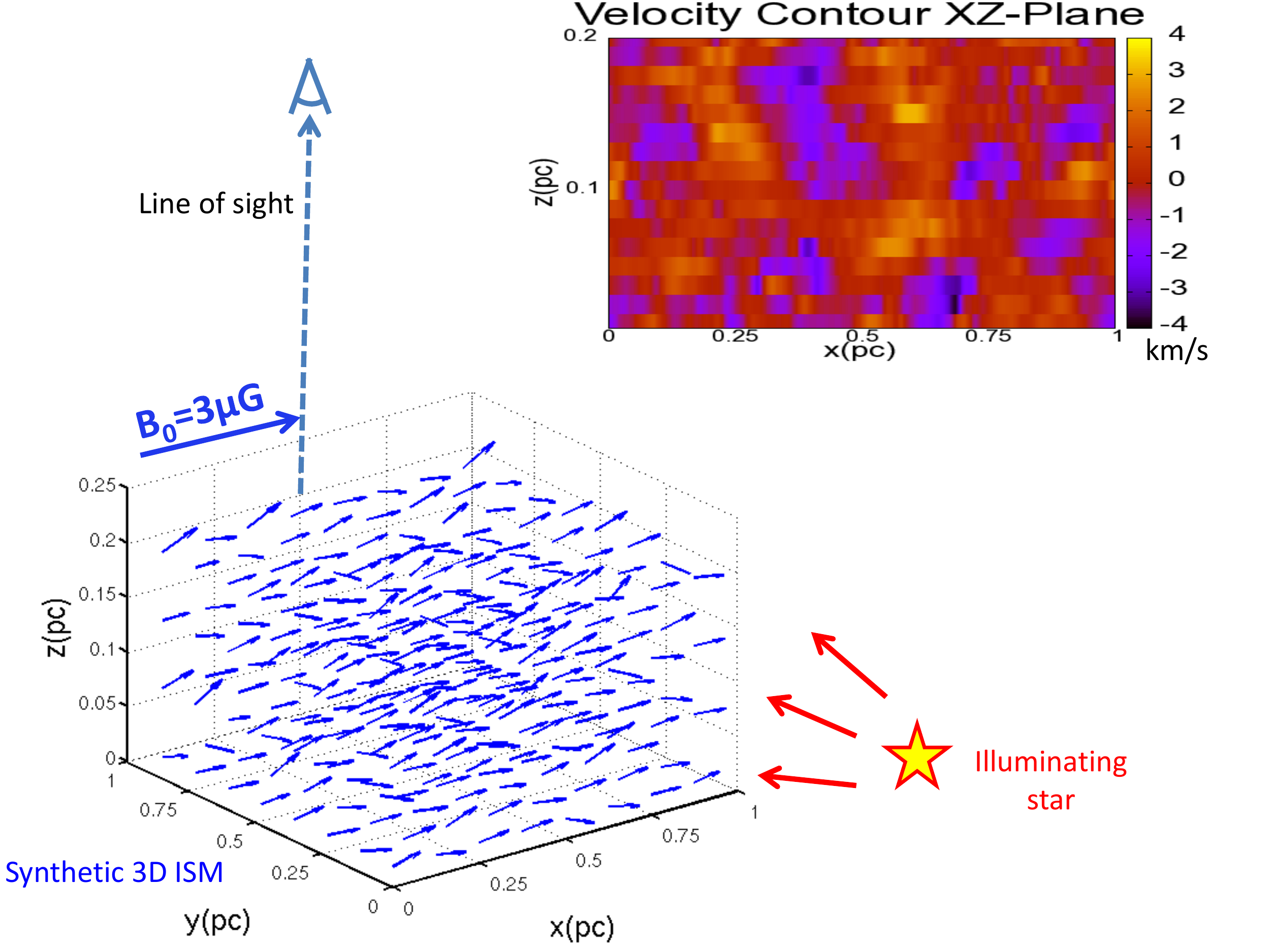}\label{H2geo}}
\subfigure[]{
\includegraphics[width=0.96\columnwidth]{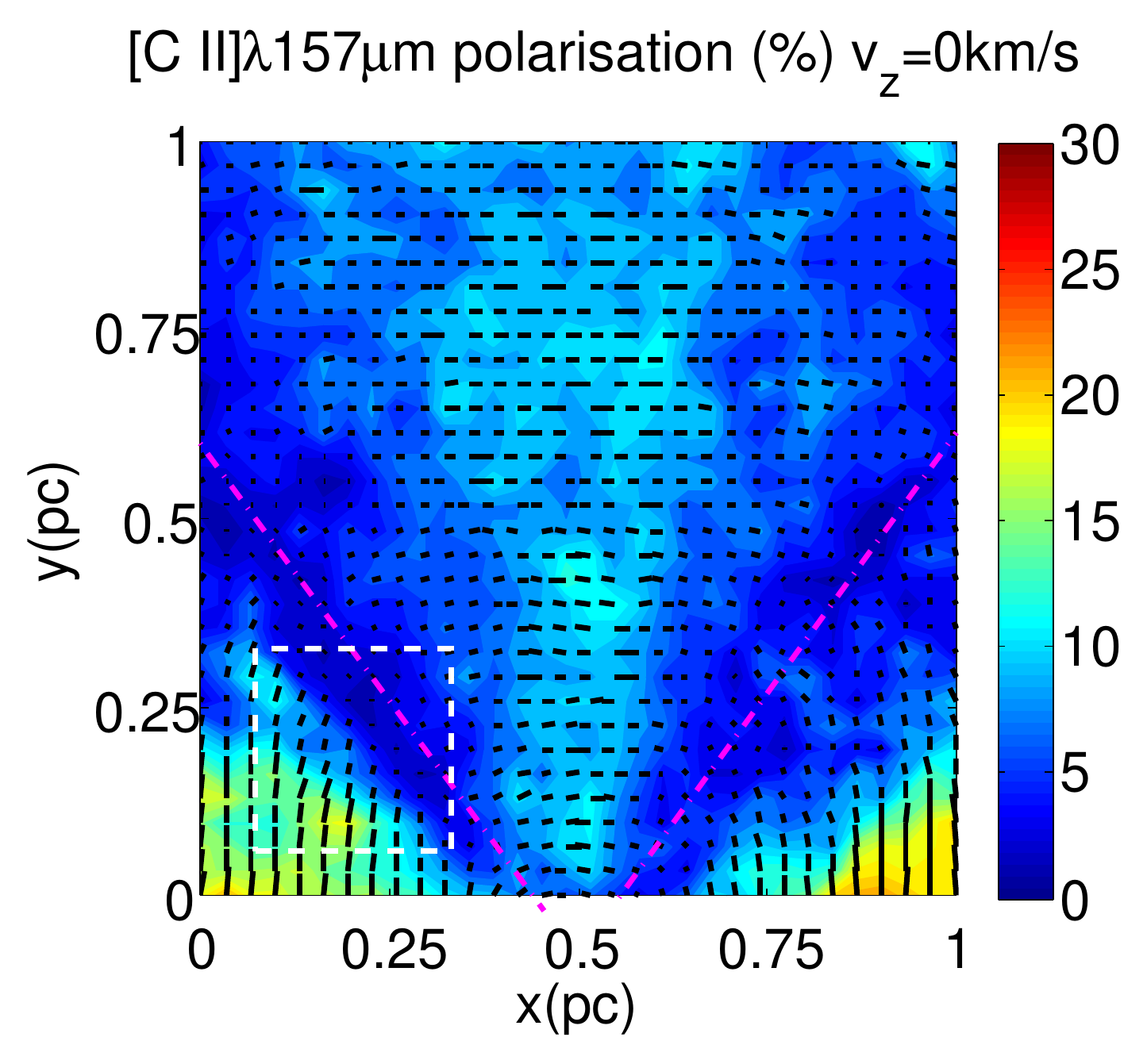}\label{polar_mapH2}}
\subfigure[]{
\includegraphics[width=0.96\columnwidth]{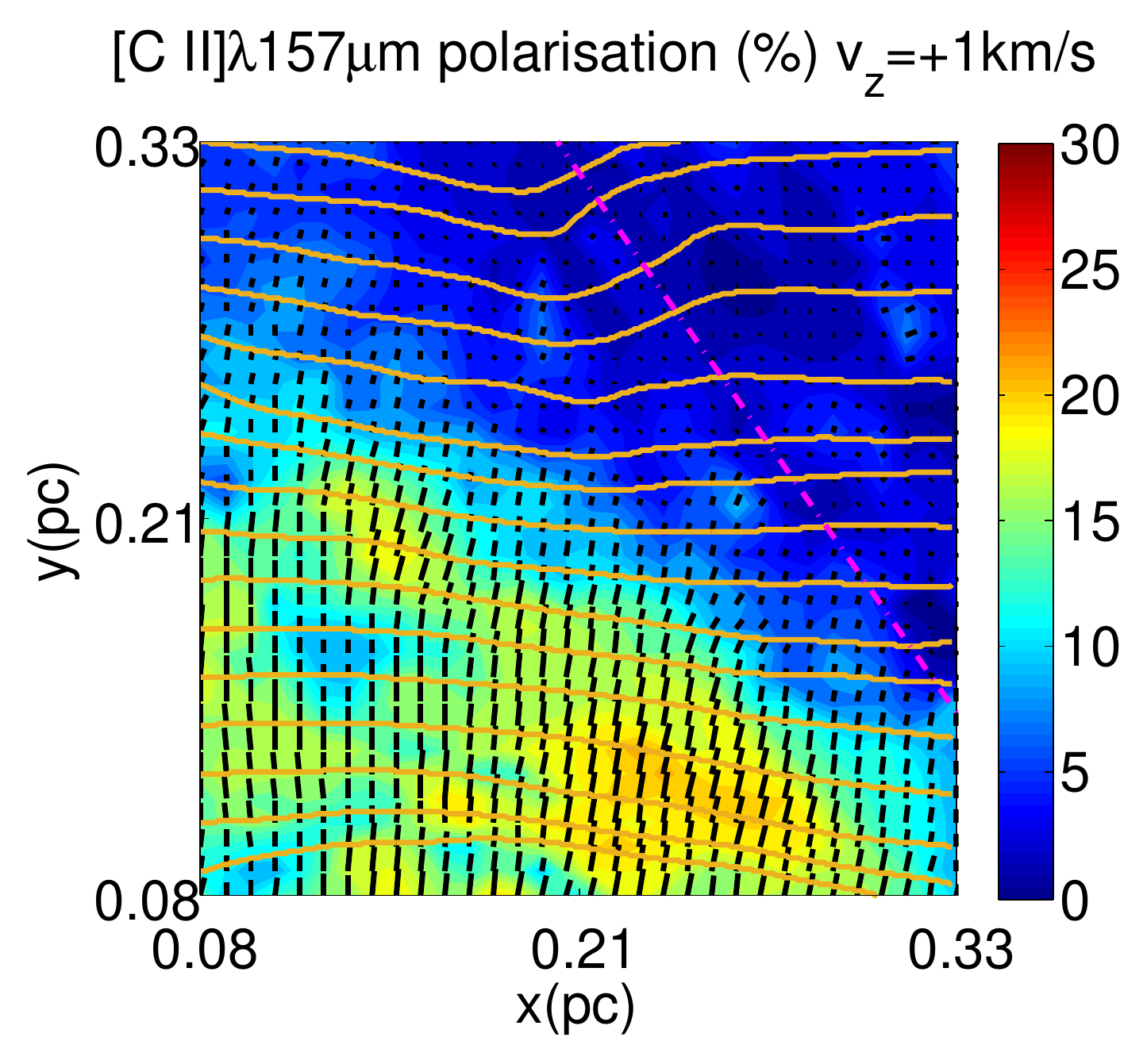}\label{polar_mapv5}}
\subfigure[]{
\includegraphics[width=0.96\columnwidth]{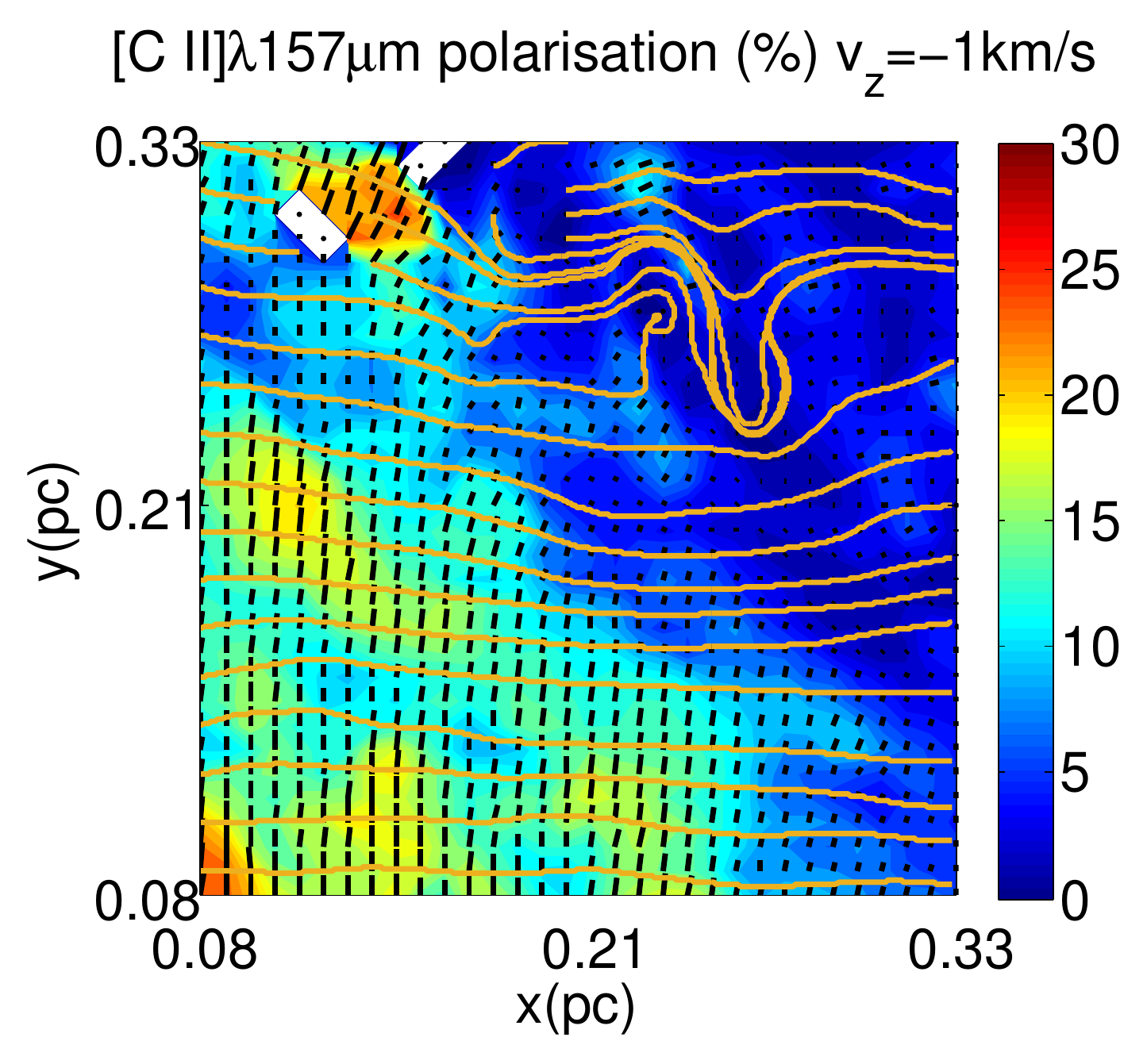}\label{polar_mapa16}}
\caption{(a)The schematic for H\,{\sc ii} Regions. The mean magnetic direction $B_0$ is parallel to the $x-$axis. Blue arrows are the magnetic direction at the corresponding grid. The radiation source is a B9-type star ($T=1\times10^4K$). Different line-of-sight velocities $v_k$ are marked with different colours in the velocity contour map in the upper right; (b) The full polarisation map cutting at $v_z=0km/s$ with the $17"$ angular resolution. Bars represent the direction of polarisation at the corresponding grid with the length of bars and the colour on the background showing the degree of polarisation. The dashed-dot purple lines denote the theoretical sign-flipping criteria for polarisation (see \S 2); (c)(d) The polarisation maps with the $4.5"$ angular resolution for the white-square region in (b) cutting at $v_z=+1km/s$ and $v_z=-1km/s$, respectively. The orange lines are the projection of magnetic field lines on the picture plane. Blank areas on the contour mean there being no cell with the cutting velocity along the line of sight in the corresponding grid.}\label{poma pH2}
\end{figure*}

On the scale equal or smaller than a few $pc$, the magnetic field can be approximated with a mean direction because the coherence length of interstellar magnetic field is in the $hpc$ scale (see \citealt{Armstrong95,CLpwl2010}). MHD simulations are performed to generate a $512\times512\times16$ trans-Alfvenic turbulence data cube with the mean magnetic field $B_0=3{\mu}G$, a typical H\,{\sc ii} Region. The Alfvenic-Mach number of the generated ISM is 1.06. A massive B9-type star radiates UV-photons to illuminate the medium. As demonstrated in \S 2, the atoms in the ISM are aligned and thus the submillimetre transitions within the ground state of atoms are polarised. Synthetic observations for the polarisation of fine-structure emission lines are performed on the simulated ISM with the $xy-$axis being the coordinates system for the polarimeter of the telescope. Presented in Fig.~\ref{H2geo} is the schematic of the synthetic ISM. The simulated area corresponds to an $1pc\times1pc$ area on the picture plane. The ISM is observed along the $z-$direction. The line-of-sight optical depth $\tau_0=0.2pc$ is sliced into 16 layers. The velocity at each grid is assumed to have a Gaussian-distribution broadening with $\sigma=v_A=0.36km/s$, in which $v_A$ is the Alfven speed obtained from the simulation. The angle between the magnetic direction at the $k$th layer and the $x-$axis is $\psi_k$ and the line-of-sight velocity at the $k$th layer is $v_k$. The line-of-sight velocity is resolved with a spectral resolution $\delta v=0.2km/s$, which is depicted by the colour of the velocity contour in Fig.~\ref{H2geo}. A line-of-sight integration at the velocity cut $v_z=v_0$ with a spectral resolution $\delta v$ is performed by selecting the grids with corresponding line-of-sight velocity $v_0-\delta v\le v_k\le v_0+\delta v$, and considering the density of the grid as the weighting parameter. The Stokes parameters at the $k$th layer are denoted by [$I_k,Q_k,U_k,V_k$]. The linear terms of the observed Stokes parameters at velocity $v_0$ are therefore:
\begin{equation}\label{polarmapin}
\begin{split}
Q(v_0)&=\int^{\tau_0}_0\int^{v_0+\delta v}_{v_0-\delta v}\mathcal{N}(v_k,\sigma^2)\rho(Q_k\cos2\psi_k+U_k\sin2\psi_k)dvd\tau,\\
U(v_0)&=\int^{\tau_0}_0\int^{v_0+\delta v}_{v_0-\delta v}\mathcal{N}(v_k,\sigma^2)\rho(-Q_k\sin2\psi_k+U_k\sin2\psi_k)dvd\tau.
\end{split}
\end{equation}

Fig.~\ref{polar_mapH2} demonstrates the full polarisation map with the $17"$ angular resolution cutting at $v_z=0 km/s$. The bars on the map depict the direction of polarisation at the corresponding grid. The length of the bars are proportional to the degree of polarisation, which is also marked by the colour on the background contour. As seen from Fig.~\ref{polar_mapH2}, the dominant polarisation direction is along either $x-$axis or $y-$axis direction. The measurability of the polarisation of the submillimetre lines is substantial because the grid points, except for those near the flipping criteria (see definition in \S 2, marked in Fig. 3b), show more than $10\%$ of polarisation. The white-squared area in Fig.~\ref{polar_mapH2} is measured with a higher resolution ($4.5"$) at different velocity cuts in Fig.~\ref{polar_mapv5} and Fig.~\ref{polar_mapa16}. The projection of the magnetic field lines on the picture plane of the corresponding velocity layer are marked as the orange lines. By comparing with Fig.~\ref{polar_mapH2}, Fig.~\ref{polar_mapv5} and Fig.~\ref{polar_mapa16} demonstrate that the polarisation of the submillimetre fine-structure lines reveal the magnetic pattern in a smaller scale with the telescope of a higher angular resolution (e.g., the fluctuation in the upper left area of Fig. 3d). In addition, the difference between Fig.~\ref{polar_mapv5} and Fig.~\ref{polar_mapa16} shows that the polarisation measured at different velocity slices can be used to depict the line-of-sight magnetic fluctuation. Analysing the polarisation of submillimetre lines with a higher spectral resolution thus gives us ample information of the interstellar turbulence on various scales.

\section{Discussion}

The numerical simulations performed in this paper demonstrate the applicability of the polarimetry of atomic lines in tracing magnetic fields in ISM. The line-of-sight dispersion of magnetic fields in real H\,{\sc ii} Regions can only be less and the degree of polarisation observed should be higher than that obtained from the synthetic observations in this paper, because the synthetic observations performed are in trans-Alfvenic ISM whereas the scale of H\,{\sc ii} Regions (normally a few $pc$) is 2 decades smaller than the injection scale of the magnetic fields in the galaxy ($100pc$, see \citealt{Armstrong95,CLpwl2010}). Moreover, the simulations performed in this work are dimensionless thus this paper also unveil the potential of studying magnetic fields with the polarimetry of atomic lines in other environments. For instance, observations of the submillimetre spectro-polarimetry from the medium in interplanetary disk will provide us the opportunity to study the local turbulence. On one hand, the spatial distribution of the interplanetary magnetic turbulence can be attained by tracing the polarimetry of atomic lines from comets, along their trajectories as demontrated in Fig.~\ref{polar_90tht0} in a way similar to Na D2 line that was studied earlier \citep{SY2013}. On the other hand, the time evolution of the turbulence in the magnetosphere of the planets can be studied by tracing the polarisation of atomic lines due to the absorption in the magnetosphere. Furthermore, the submillimetre fine-structure lines such as C\,{\sc ii}$158\mu m$ from PDRs are bright in those early star-forming galaxies (see, e.g., \citealt{Lagache2017C2}). The neighbouring stars and the AGN (Active Galactic Nuclei) in these galaxies can provide anisotropic radiation fields and thus those fine-structure lines can be polarised according to the direction of magnetic fields in the early galaxies. Last but not the least, the multi-species atomic and ionic lines detected from the illuminated medium between galaxies- such as in Magellanic Stream \citep{FoxRichter2013Magstream} and the H\,{\sc i} envelope around the NGC$4490/4485$ system \citep{Clemens1998envelope}- can be utilised to investigate the intergalactic magnetic fields with atomic alignment.

\section{Conclusion}

In summary, we have, for the first time, applied numerical simulations to study the polarisation of submillimetre fine-structure lines in the ISM. The main aims of our simulations are to evaluate the measurability of the polarisation of submillimetre atomic lines and to investigate what information of the interstellar magnetic field it can provide. The MHD simulations are carried out to generate the diffuse ISM turbulence. The synthetic observations of the submillimetre atomic lines are performed for the turbulent ISM. The polarisation maps produced from the observations are compared with the magnetic fields in the ISM. We should emphasise that all the conclusions from the simulations throughout the paper are not limited to the example [C\,{\sc ii}]$\lambda157\mu m$ emission line but generically applicable to all the submillimetre absorption and emission atomic lines. Our main conclusions are:

\begin{itemize}
\item Polarisations in submillimetre absorption and emission atomic lines are detectable from the diffuse ISM due to the alignment from optical/UV excitations.
\item The direction of polarisation of submillimetre atomic lines reveals the 2D magnetic field in the plane of sky with a $90^{\circ}$-degeneracy; 3D magnetic fields can be detected if polarisations of multiple ($\ge2$) submillimetre lines are detected.
\item The degree of polarisation is affected by the temperature of the pumping star.
\item The polarisation of atomic lines with resolved spectrum reveals the magnetic fluctuation along the line of sight and thus is a promising tool to study interstellar turbulence.
\item Submillimetre spectropolarimetry is a valuable magnetic tracer that provides us with multi-scale magnetic patterns within and beyond our galaxy, from interplanetary medium to the early universe.
\end{itemize}

\section*{Acknowledgements}

We are grateful to Helmut Wiesemeyer and Alexey Chepurnov for the discussions on the synthetic observations. We have also benefited from fruitful discussions with Michael Vorster and Reinaldo Santos de Lima on the simulations of interstellar environment. Finally, We thank the referee, Alexandre Lazarian, for valuable comments and suggestions.



\bibliographystyle{mnras}
\bibliography{Zhan0306} 



%
%
%


\bsp	
\label{lastpage}
\end{document}